\documentclass[12pt]{iopart}

\usepackage{amssymb}
\usepackage{graphicx}
\usepackage{bm}
\usepackage{amsfonts}

\begin{document}

\title{Optimal control of a charge qubit in a double quantum dot with a Coulomb impurity}

\author{Diego S. Acosta Coden, Rodolfo H. Romero, Alejandro Ferr\'on, and Sergio S. Gomez}
\address{Instituto de Modelado e Innovaci\'on Tecnol\'ogica (CONICET-UNNE) \\ and Facultad de Ciencias Exactas y Naturales y Agrimensura, Universidad Nacional del Nordeste, Avenida Libertad 5400, W3404AAS Corrientes, Argentina}

\date{\today}


\begin{abstract}
We study the efficiency of modulated laser pulses to produce efficient
and fast charge localization transitions in a two-electron double
quantum dot. We use a configuration interaction method to calculate
the electronic structure of a quantum dot model within the effective
mass approximation. The interaction with the electric field of the
laser is considered within the dipole approximation and optimal
control theory is applied to design high-fidelity ultrafast pulses in
pristine samples. We assessed the influence of the presence of Coulomb
charged impurities on the efficiency and speed of the pulses. A
protocol based on a two-step optimization is proposed for preserving
both advantages of the original pulse. The processes affecting the
charge localization is explained from the dipole transitions of the
lowest lying two-electron states, as described by a discrete model
with an effective electron-electron interaction.

\end{abstract}
\maketitle
\section{Introduction}

Semiconductor quantum dots (QDs) are excellent candidates for realizing
qubits for quantum information processing because of the manipulability, 
scalability and tunability of their electronic and optical
properties \cite{Loss1998,DiVincenzo2005,Petta2005,Brunner2011}. Advances in 
semiconductor technology allow the preparation of 
complex structures and a fine experimental control of the parameters defining 
their electrical and optical properties \cite{QD1,QD2,QD3}. 

In recent years there has been an increasing interest in controlling
quantum phenomena in molecular systems and nanodevices \cite{oct1,oct2,oct3,mur,fosjap,acosta1,Rasanen07,Rasanen2008}, due to the 
possibility to modify the wave function of the 
system through the appropriate tailoring of external fields such as laser 
pulses. 
Coherent quantum control of electrons in quantum dots exposed to 
electromagnetic radiation is of great interest in many technological 
applications from charge transport devices to quantum
information \cite{oct7,oct8,oct9}. Several studies in quantum control of 
double quantum dots (DQDs) has been performed using
gate voltages and optimized laser pulses \cite{oct10,oct11,oct12}. 
In addition, the number of quantum control experiments is rapidly 
rising through the improvement of laser pulse shaping and closed-loop
learning techniques\cite{oct10,oct11,oct12}. 

Among other techniques, Optimal Control Theory (OCT) \cite{oct13,oct14,oct15,oct16} has become
an efficient tool for designing laser pulses able to control
quantum processes. The optimal field is the field employed in order
to steer a dynamical system from a initial state
to a desired target state minimizing a cost functional
which generally penalizes the energy (fluence) of the
pulse. A great effort has been invested in recent years in the development of different methods
in order to solve the optimal equations \cite{Sundermann99, Zhu98, Maday03, Ohtsuki04, moct1}. Monotonically convergent iterative schemes proposed by 
Tannor et al. \cite{oct17} and Rabitz et al. \cite{oct17b}
have been successfully applied to the control of
different quantum phenomena, mainly related to chemical process \cite{oct18,oct19}. In the last years, optimal control theory became a research area that 
has received increasing interest from the scientists studying emerging 
fields within quantum information science \cite{loss,nep1}. Modern quantum devices are 
systems where the wave function must be manipulated with highest possible 
precision using, for example, quantum gates. This high-fidelity quantum 
engineering needs new and efficient strategies which allow an optimal 
suppression of environment losses during gating or optical control.  
In addition, quantum computation theoretically requires extremely high 
fidelity in the elementary quantum state transformations. Optical control of 
quantum dot based devices is of fundamental interest for a wide range of 
applications in quantum information \cite{loss,nep1,mod1,Werschnik05,Werschnik07,Rasanen12}. For example, optical manipulation is an 
alternative in order to store qubits in the electron spin \cite{loss}. 
Hansen {\it et. al} \cite{nep1} show that carefully selected microwave pulses 
can be used to populate a single state of the first excitation band in a 
two-electron DQD and that the transition time can be decreased using optimal 
pulse control. Heller {\it et. al} \cite{rans1} show that the quality 
of periodic recurrence (quantum revival) in the time evolution in a quantum 
well, can be restored almost completely by coupling the system to an 
electromagnetic field obtained using quantum optimal control theory.
It results clear that the study of quantum dynamics of nanodevices and
the possibility of controlling different processes in such systems represent
an important research field with very interesting technological applications.
The implementation of numerical techniques such as OCT allow us to analyze
different possibilities and scenarios in order to construct and improve
nanodevices based quantum bits.

It is known that the presence of impurity centers has a great influence on 
the optical and electronic properties of nanostructured
materials. Recent works \cite{dasarma1,coden,bast,Ashoori92,im1,im2,ex5} studied the effects of having 
unintentional charged impurities in two-electron laterally coupled 
two-dimensional double quantum-dot systems. They analyzed the effects of 
quenched random-charged impurities on the singlet-triplet exchange coupling 
and spatial entanglement  in two-electron double quantum-dots. Although
there is an enormous interest in applying these systems in quantum
information technologies, there are few works trying to quantify the effect 
of charged impurities on this kind of tasks. The existence of unintentional 
impurities, which are always present in nanostructured devices, affects 
seriously the possibility of using these devices as quantum bits. Although 
the distribution and concentration of impurities in these systems result 
unknown parameters, there are some recent works that propose the possibility
of experimentally control these issues \cite{ex1,ex4,ex2,ex3}. Impurity 
doping in semiconductor materials is considered as a useful technology that 
has been exploited to control optical and electronic properties in different 
nanodevices. 

It is worth to mention that, due to environmental perturbations, these systems 
lose coherence. For example, confined electrons interact with spin nuclei 
through the hyperfine interaction leading, inevitably, to decoherence 
\cite{Petta2005}. Even, having just one  charged impurity could induce qubit 
decoherence if this impurity is dynamic and has a fluctuation time scale 
comparable to gate operation time scales \cite{dasarma1}. Decoherence is a 
phenomenon that plays a central role in quantum information and its 
technological applications \cite{dec1,dec2,dec3,dec4,dec5,dec6,dec7,nm1,nm2,nm3}. The
short transition time in optically driven processes reduces the effect of 
decoherence sources, such as hyperfine or phonon interactions. A number of 
known quantum control techniques such as quantum bang-bang control 
\cite{oct22} or spin-echo pulses \cite{Petta2005} allow experimentalists to 
fight decoherence. It seems reasonable that optimal control theory can be 
considered as a tool in order to design control pulses or gates so that quantum
systems can be controlled in presence of environmental couplings without 
suffering significant decoherence.

The aim of this work is to present a detailed analysis of the optical control
of two electrons in a two-dimensional coupled quantum dot and the effect of 
impurities by means of Optimal Control Theory. 
The paper is organized as follows. In Sec. \ref{DQD} we introduce the model 
for the two-dimensional two-electron coupled 
quantum dot and briefly describe the method used to calculate its electronic 
structure. In Sec. \ref{oct}, we describe optimal control equations for a 
two-dimensional two-electron coupled quantum dot. In Sec. \ref{cont} we analize
the operation of a charge qubit in presence of impurities.
In Sec. \ref{iniop}, we propose a protocol of initialization (Sec. \ref{ini}) 
and  operation (Sec. \ref{op}) of the qubit using electromagnetic pulses 
avoiding the effects of the impurities by means of OCT. Finally, 
In Sec. \ref{con} we summarize the conclusions with a discussion of the most 
relevant points of our analysis.

\section{Model and calculation method}

\label{DQD}

We consider two laterally coupled two-dimensional quantum dots whose 
centers are separated a distance $d$ from each other, and containing two 
electrons. In quantum dots electrostatically produced, both their size and 
separation can be controlled by variable gate voltages through metallic 
electrodes deposited on the heterostructure interface. The eventual existence 
of doping hydrogenic impurities, probably arising from Si dopant atoms in the 
GaAs quantum well, have been experimentally studied \cite{Ashoori92}. These 
impurities have been theoretically analyzed with a superimposed attractive 
$1/ r$-type potential \cite{im1,im2}.
Furthermore, some avoided crossing and lifted degeneracies in the spectra 
of single-electron transport experiments have been attributed to negatively 
charged Coulomb impurities located near to the QD \cite{ex5}. From fitting 
the experimental transport spectra to a single-electron model of softened 
parabolic confinement with a Coulomb charge $q$, a set of parameters are 
obtained; among them, a radius of confinement of $15.5$ nm, a confinement 
frequency $\hbar\omega=13.8$ meV and an impurity charge of approximately 
1 or 2 electron charges. Indeed, the 
uncertainty in the parameters and the suppositions introduced in the model 
does not allow one to precisely ensure the impurity charge, with the 
screening probably reducing its effective value to less than an electron 
charge. Therefore, we consider the charge of the doping atom $Ze$ as a 
parameter varying in the range $0\leq Z \leq 1$, in order to explore its 
effect on the properties of the system.
\subsection{Electronic structure and dynamics}
In this work we model the Hamiltonian of the two-dimensional two-electron 
coupled quantum dot in presence of charged impurities within the single 
conduction-band effective-mass approximation \cite{mod1}, namely,

\begin{equation}
\label{h1}
H_0=h({\bf r}_1)+h({\bf r}_2)+\frac{e^2}{4\pi\varepsilon \varepsilon_0 r_{12}},
\label{Hamiltonian}
\end{equation}

\noindent where ${\bf r}_i=(x_i,y_i)$ ($i=1,2$) and

\begin{equation}
\label{h2}
h({\bf r})=-\frac{\hbar^2}{2m^*}\nabla^2+V_L({\bf r})+V_R({\bf r})+V_A({\bf r}),
\end{equation}

\noindent where $h({\bf r})$ is the single-electron Hamiltonian that includes 
the kinetic energy of the electrons, in terms of their effective mass 
$m^*$, and the confining potential for the left and right quantum dots $V_L$ 
and $V_R$, and the interaction of the electrons with the charged impurities, 
$V_A$. 

The last term of the Hamiltonian, Eq. (\ref{Hamiltonian}), represents the 
Coulomb repulsive interaction between both electrons at a distance 
$r_{12}=|{\bf r}_2-{\bf r}_1|$ apart from each other, within a material of 
effective dielectric constant $\varepsilon$.
We model the confinement with Gaussian attractive potentials

\begin{equation}\label{h3}
V_{i}({\bf r})=-V_0\exp\left(-\frac{1}{2a^2} |{\bf r}-{\bf R}_{i}|^2\right), \ (i=L, R),
\end{equation}

\noindent where ${\bf R}_L$ and ${\bf R}_R$ are the positions of the center 
of the left and right dots, $V_0$ denotes the depth of the potential and $a$ 
can be taken as a measure of its range.
Along this work, we will consider a single impurity atom centered 
at ${\bf R}_A$, and modelled as a hydrogenic two-dimensional Coulomb potential

\begin{equation}\label{h4}
V_A({\bf r})= -\frac{Ze^2}{4\pi\varepsilon \varepsilon_0 
|{\bf R}_A-{\bf r}|}
\end{equation}

Since the Hamiltonian does not depend on the electron spin, its eigenstates 
can be factored out as a product of a spatial and a spin part

\begin{equation}
\Psi_i({\bf r}_1,{\bf r}_2,m_{s_1},m_{s_2}) = \Psi_i^{S}({\bf r}_1,{\bf r}_2) \chi_{S,M},
\end{equation}

\noindent where $S=0$, 1 for singlet and triplet states, respectively, 
and $M = m_{s_1} + m_{s_2}$ is the total spin projection.

The eigenstates of the model Hamiltonian can be obtained by direct 
diagonalization in a finite basis set \cite{var1}. 
The spatial part is obtained, in a full configuration 
interaction (CI) calculation, as 

\begin{equation}\label{variational-functions}
\Psi_m^S({\bf r}_1,{\bf r}_2) = \sum_{n=1}^{N_{\rm conf}} c^S_{mn} \Phi^S_{n}({\bf r}_1,{\bf r}_2)
\end{equation}

\noindent where $N_{\rm conf}$ is the number of singlet ($S=0$) or triplet 
($S=1$) two-electron configurations $\Phi^S_{n}({\bf r}_1,{\bf r}_2)$ 
considered, and $n=(i,j)$ is a configuration label obtained from the indices 
$i$ and $j$ from a single electron basis, i.e.,

\begin{equation}
\Phi^S_{n}({\bf r}_1,{\bf r}_2) = \frac{1}{\sqrt{2}}\left[ \phi_i({\bf r}_1)\phi_j({\bf r}_2)+(1-2S)\phi_j({\bf r}_1)\phi_i({\bf r}_2)\right]
\end{equation}

\noindent for $i\ne j$, and $\Phi^{S=0}_{n}({\bf r}_1,{\bf r}_2)=
\phi_i({\bf r}_1)\phi_i({\bf r}_2)$ for the doubly occupied singlet states.

We chose a single-particle basis of Gaussian functions, centered at the dots 
and atom positions ${\bf R}_P$ ($P = L, R, A$), of the type \cite{sr1,sr2}

\begin{equation}\label{variational-functions}
\phi_{i}({\bf r})=N (x-P_x)^{m_i} (y-P_y)^{n_i} \exp\left(-\alpha_i|{\bf r}-{\bf R}_P|^2\right), 
\end{equation}

\noindent where $N$ is a normalization constant, and $\ell_i=m_i+n_i$ 
is the $z$-projection of the angular momentum of the basis function. 
The exponents $\alpha_i$ were optimized for a single Gaussian well and 
a single atom separately, and supplemented with extra functions when used 
together.
For our calculations a basis set of $2s 2p$ functions for the dots, 
and $5s 5p 1d 1f$ for the atom was found to achieve converged results for 
the energy spectrum.  

The numerical results presented in this work refers to those corresponding 
to the parameters of GaAs: effective mass $m^*=0.067 m_e$, effective 
dielectric constant $\varepsilon=13.1$, Bohr radius $a^*_B=10$ nm and 
effective atomic unit of energy 1 Hartree$^*=10.6$ meV 
\cite{ex5,dasarma1}. 

The depth of the Gaussian potentials modelling the dots 
are taken as $V_0=4$ Hartree$^*=42.4$ meV, and its typical range 
$a=\sqrt{2}a_B^*=14.1$ nm, with an interdot separation 22.5 nm.
Smaller interdot separations provides a  high electric dipole moment and, hence, strong coupling with the laser field but induce delocalization of the electrons making difficult to to define the occupation on a single dot. Larger interdot separation produces the opposite effect, with the drawback of small coupling with the laser electric field, thus worsening the controlability of the QDs. 
\subsection{Optimal Control Theory for electrons interacting with a time-dependent electric field}
\label{oct}

Let us consider a laser field  ${\bf\varepsilon}(t)$ pointing along the line joining the QDs ($x$ direction) and propagating along the $z$ direction (perpendicular to the plane of the system). The time evolution of the electron state, assuming the dipole approximation for the interaction, will be given by
\begin{eqnarray}
i\frac{\partial \Psi(t)}{\partial t}&=&H
\Psi(t)\\
H&=&H_0-\mu{\varepsilon}(t)
\end{eqnarray}
where $\mu$ is $x$-component of the dipole moment operator.

Application of optimal control theory (OCT) allows to design a pulse of duration $T$, whose interaction drives the system to a state $\Psi(T)$, having maximum overlap to a given state $\phi_F$ or, equivalently, maximizes the expectation value of the operator ${\cal O}=|\phi_F\rangle\langle\phi_F|$ at the end of the pulse application \cite{oct16}: 

\begin{equation}
J_1[\Psi]=|\langle\Psi(T)|\phi_F\rangle|^2
\end{equation}

\noindent where $J_1$ is known as the yield. In order to avoid high energy fields we introduce
a second functional

\begin{equation}
J_2[{\bf\varepsilon}]=-\alpha\left[\frac{1}{T}\int_0^T{\bf\varepsilon}^2(t)dt-
F \right]
\end{equation}

\noindent where the time-integrated intensity is known as the fluence of the field, $F$ is the fixed fluence
and $\alpha$  is a time-independent Lagrange multiplier. In addition the electronic wave function has to 
satisfy the time-dependent Schr\"odinger equation, introducing a third 
functional:

\begin{equation}
J_3[{\bf\varepsilon},\Psi,\chi]=-2Im \int_0^T \left\langle\chi(t)\left|\frac{\partial}{\partial t}-H(t)\right
|\Psi(t)\right\rangle
\end{equation}

\noindent where we have introduced the time-dependent Lagrange multiplier $\chi(t)$. Finally, the Lagrange functional has the form $J=J_1+J_2+J_3$. The Variation of this functional with respect to 
$\Psi(t)$, ${\bf\varepsilon}(t)$ and $\chi(t)$ allows us to 
obtain the control equations \cite{moct1}

\begin{eqnarray}
i\frac{\partial \Psi(t)}{\partial t}&=&H
\Psi(t)\;,\; \Psi(t=0)=\phi({\bf r}_1,{\bf r}_2)\\
\frac{\partial \chi(t)}{\partial t}&=&H(t)\chi(t),\chi(T)=|\phi_F\rangle\langle\phi_F|\Psi(T)\rangle\\
{\bf\varepsilon}(t)&=&-\frac{1}{\alpha}Im\langle\chi(t)|\hat{\mu}|\Psi(t)\rangle\\
\int_0^T{\bf\varepsilon}^2(t)dt&=&A_0
\end{eqnarray}

\noindent This set of coupled equations can be solved iteratively, for example, using the efficient 
forward-backward propagation scheme developed in \cite{oct17}. The algorithm starts by 
propagating $\phi({\bf r}_1,{\bf r}_2)$ forward in time, using in the first step a guess of the laser 
field ${\bf\varepsilon}^0(t)$. At the end of this step we obtain the
wave function $\Psi^{(0)}({\bf r}_1,{\bf r}_2,T )$, which is used to evaluate
$\chi^{(0)}({\bf r}_1,{\bf r}_2,T)=|\phi_F\rangle\langle\phi_F|\Psi^{(0)}({\bf r}_1,{\bf r}_2,T )\rangle$.
The algorithm  continues with propagating $\chi^{(0)}({\bf r}_1,{\bf r}_2,t)$ backwards in time. 
In this step we need to know both wave functions ($\Psi^{(0)}$ and $\chi^{(0)}$) at the same time. In this step we obtain the first optimized pulse ${\bf\varepsilon}^1(t)$. We repeat this operation until
the convergence of $J$ is achieved. The Lagrange multiplier $\alpha$ is calculated using the fixed fluence following the calculation details showed in reference\cite{oct17}. The numerical integration of the forward and backward time evolution was performed using fourth-order Runge-Kutta algorithms.

As usual in this technique, we constrain the field using an envelope function 
\cite{moct1,Sundermann99},
\begin{equation}
f(t)=\frac{1}{2}\left\{{\rm erf}\left[\frac{a}{T}\left(t-\frac{T}{b}\right)\right]+
{\rm erf}\left[\frac{-a}{T}\left(t-T+\frac{T}{b}\right)\right]\right\}
\end{equation}
in order to have a electromagnetic pulse ${\bf\varepsilon}(t)$ with a finite duration, i.e., ${\bf\varepsilon}(0)={\bf\varepsilon}(T)=0$. Furthermore, a spectral cut-off is applied to remove frequency components higher than a prescribed threshold $\omega_c$ \cite{oct17}.
\section{Controllability of the charge qubit states in the presence of an 
impurity}
\label{cont}
\begin{figure}[ht]\centering
\includegraphics[scale=0.5]{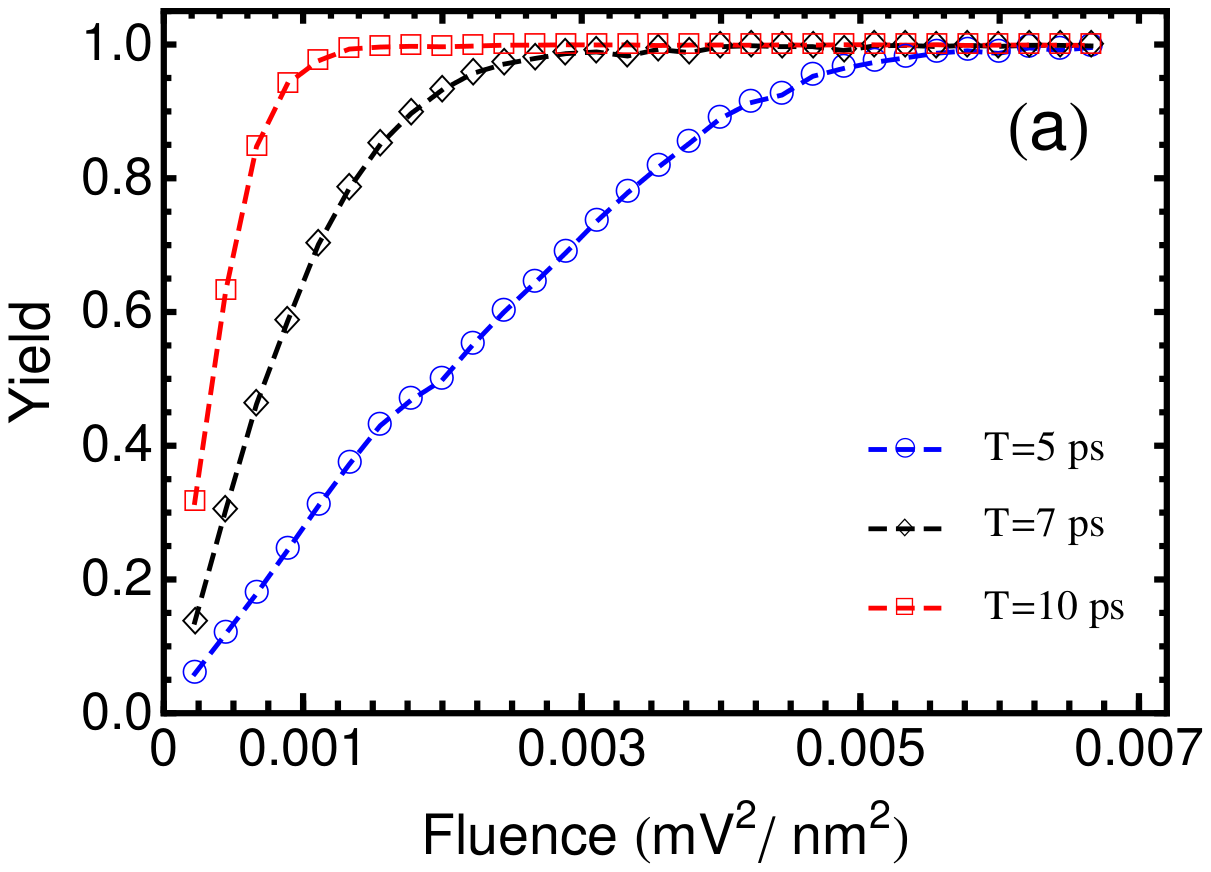} \\
\includegraphics[scale=0.5]{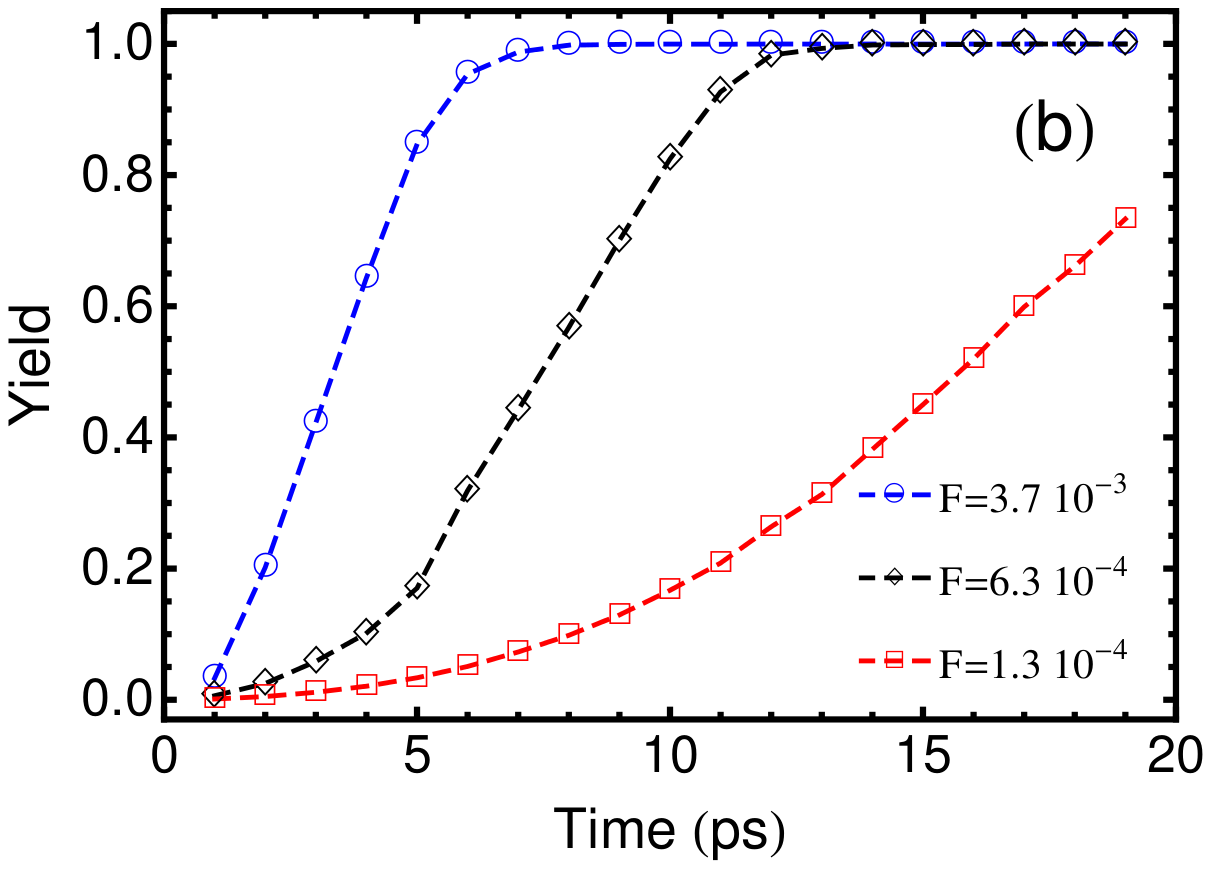} \\

\caption{\label{Yield-Fluence-Time} (Color online) (a) Yield as a function of 
fluence $F$, for three different values of pulse duration $T=5$, $7$ and $10$ 
ps. (b) Yield as a function of time duration of the pulse for three different 
values of fluence, $F=3.7\times 10^{-3}$, $6.3\times 10^{-2}$ and $1.3\times10^{-3}$ mV$^2$/nm$^2$ps.}
\end{figure}

Firstly, we will analyze the controllability of a symmetrical DQD in a clean sample. 
The absence of impurities means that the only relevant potentials are those from the confining wells. 
Then, the electronic ground state, $\Psi_0({\bf r}_1,{\bf r}_2)$, is spatially delocalized, with a high 
density around each dot. The first and second excited states, $\Psi_1$ 
and $\Psi_2$, have very close energies, what allows one to define the doubly 
occupied states at the left and right dots as 
$\Psi_{LL/RR}({\bf r}_1,{\bf r}_2)=2^{-1/2}\left[\Psi_1({\bf r}_1,{\bf r}_2) 
\pm \Psi_2({\bf r}_1,{\bf r}_2)\right]$. Let us consider one of them, say 
$\Psi_{\rm RR}$, as the target state for optimizing the laser field. Thus the 
optimal field will produce a delocalization-localization transition of the 
electron charge, which could be detected by measuring the charge variation at 
the dots. Fig. \ref{Yield-Fluence-Time} shows the yield obtained in the 
transition $\Psi_0\rightarrow\Psi_{\rm RR}$ where one electron is moved from the 
left to the right dot.  As we can observe in the lower panel of 
Fig. \ref{Yield-Fluence-Time}, for a fixed fluence, a high yield is obtained 
by extending the pulse duration. On the other hand, for a given duration of 
the pulse, the yield becomes larger as the fluence increases (as shown in
upper panel of Fig. \ref{Yield-Fluence-Time}). Therefore, 
the most convenient situation corresponds to that where a long pulse of high 
intensity is applied. However, the pulse duration cannot be arbitrarily long 
because the coherence time have the order of nanoseconds. So, in 
order to perform operations cyclically with the qubit, the pulse has to be restricted to 
tens or hundreds of picoseconds.

Now consider the effect produced by charged impurities. If the sample were heavily doped, that is, it has a high impurity density or their charges are high (i.e., comparable to the electron charge) the system is far from being controllable. Therefore, we shall consider the situation where the density of impurities is low, such that no more than one of them is in the neighbourhood of the DQD. We also assume that their charge $Ze$  is small, characterized by an effective parameter $Z<0.5$ (meaning that it is highly screened), a range which has been considered suitable in a previous work \cite{sr1}. 

\begin{figure}\centering
\includegraphics[scale=0.6]{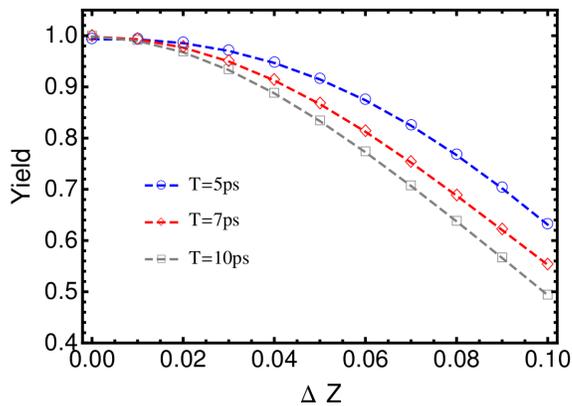}
\caption{\label{yield-Z} (Color online)  Calculated yields for pulses of $5$, 
$7$ and $10$ ps as a function of the magnitude of the effective charge $Z$ of 
an impurity localized in the middle of the dots. The pulses were optimized for 
the double quantum dot without impurity, with a fluence of 
$F =1.3\times10^{-3}$ mV$^2$/nm$^2$, and applied to the system containing 
the Coulomb impurity. The yield is nearly $1$ for $Z=0$, but it becomes 
strongly deteriorated even for small values of $Z$.}
\end{figure}

Fig. \ref{yield-Z} shows the calculated yield obtained when the pulse 
optimized for the transition $\Psi_0\rightarrow \Psi_{\rm RR}$ in the clean DQD, is applied to the doped DQD having a Coulomb point charge $Ze$ located at the mid point between both quantum dots. 
Fig. \ref{yield-Z} shows that for ultrashort pulses of a few picoseconds, 
even very small impurity charges can heavily deteriorate the performance of 
the optimal pulse. The unintentional impurities act as trapping centers for the electrons, spoiling the fast operation of the pulses optimally designed for the clean device.

In the next section we propose and discuss a protocol for controlling the electronic states using OCT in order to avoid the deterioration introduced by a charge impurity located in between the dots.

\section{OCT-based two-step protocol}
\label{iniop}

Although the OCT pulses produce fast transitions between localized and delocalized states in a clean 
DQD, with high fidelity, they fail after adding a Coulomb charge. This is due to the fact that, in the presence of the charge, the spatial distribution of the electron wave function of the system is not only localized around the QDs, but also in the proximity of the center of the Coulomb potential. 
Nevertheless, for a proper operation and detection of the electron charges in the QDs (e.g., using quantum point contacts) it is desirable that both the initial and target qubit states correspond to electron densities predominantly localized around the QDs.

Our proposed protocol consists in using OCT for tailoring two pulses, to be sequentially applied, in order to induce the transitions $\Psi_0\rightarrow \Psi_0^{(0)}\rightarrow \Psi_{\rm RR}^{(0)}$. 
Firstly, an {\em initialization} pulse is designed for the transition $\Psi_0\rightarrow \Psi_0^{(0)}$, i.e., a pulse by which the electrons initially in the ground state of the Hamiltonian with the impurity ($H^{(Z)}$) are set in the ground state of the Hamiltonian without impurity ($H^{(0)}$). 
Immediately afterwards, the second pulse drives the wave function to perform the same transition as in the clean system, i.e., $\Psi_0^{(0)}\rightarrow \Psi_{\rm RR}^{(0)}$.
This second pulse has all the previously discussed advantages of fast and high-fidelity evolution, and correspond to the desired {\em qubit operation}, e.g., for information processing. 

The influence of the impurity charge could, in principle, spoil and slow down the whole process because it enters in two ways:
 (i) the initialization pulse introduces an additional evolution time to set the wave function in the state $\Psi_0^{(0)}$. As a consequence, the operation of the device with an impurity, performed between 
the same pair of states, will be necessarily slower than without impurities 
present, and the fidelity, for a fixed pulse duration, could depend strongly on $Z$; 
(ii)  although the operation pulse produces a transition between impurity-free states $\Psi_0^{(0)}$ and $\Psi_{\rm RR}^{(0)}$, they are to be represented in terms of states of the whole Hamiltonian as $\Psi_i^{(0)}=\sum_j c_{ij}(Z) \Psi_j$, with coefficients depending on the charge $Z$. We shall show in the following, however, that none of these circumstances eliminates the advantages of the procedure, which holds a high fidelity with operation times lesser than the typical dephasing times.

\subsection{Initialization of the qubit}
\label{ini}

We consider three different values for the charge of the impurity, namely, $Z=0.1$, $0.2$ and $0.3$, corresponding to the conditions of weak and intermediate strength of the Coulomb potential competing with the confining ones in the QDs. 

Firstly, we calculated pulses by optimizing them without imposing any restriction on the maximum allowed frequency.
 As a result, the yield increases monotonically with the fluence until reaching a plateau 
having a maximum value of 99.9\% for $Z=0.1$ and 99.4\% for $Z=0.3$. These maximum 
yields are reached nearly at $F=1.3\times10^{-3}$ mV$^2$/nm$^2$.
On the other hand, when a cut-off frequency $\omega_c$ is imposed, the increase in the yield is similar until the aforementioned value of fluence. Nevertheless, instead of the plateaus of the 
unconstrained pulses, the yield of frequency-constrained pulses reaches a top
and then oscillates, with a slight decrease in average.
Remarkably, the use of a cut-off frequency $\omega_c=20$ THz, compatible with 
current experimental capabilities, affects more strongly the yield of the systems with smaller impurity charges. 
This behaviour can be attributed to the fact that, for low fluence pulses, the 
electron can only be promoted to the low-lying levels having small excitation 
energy. On the other hand, pulses of larger fluence entail a larger amplitude 
and excitation energies to higher states. While the unconstrained pulses have 
a suitable frequency composition to produce the excitation to the higher 
levels, the frequency constrained pulses do not have the higher frequency 
required to excite the electrons to the higher levels. Therefore, some 
transitions involving highly excited states are inhibited, leading to a 
decrease in the yield.
In other words, for systems with a large impurity charge, its electronic 
structure is mainly determined by the lowest lying (i.e., by the low frequency 
or low excitation energy) states of the Coulomb potential. Therefore, in such 
cases, both pulses, with and without cut-off, are almost equally well suited 
for producing the maximum yield which, nevertheless, becomes smaller than for 
small impurity charges. 

\begin{figure}\centering
\includegraphics[scale=0.6]{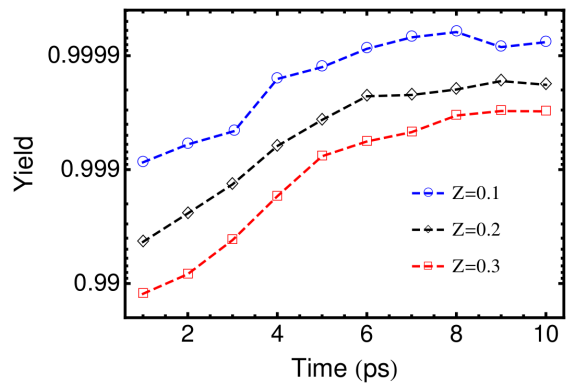}
\caption{Yield of the initialization pulse of fluence $F=1.3\times10^{-4}$ mV$^2$/nm$^2$ as a 
function of time for 
the double quantum dot with a Coulomb charge $Z$ placed in the middle of 
the segment of line joining the centers of the dots. Application of this pulse 
prepares the system to be used as a qubit, evolving the actual ground state 
of the system with an impurity, $\Psi_0$, to the ground state of the system 
having no impurity, $\Psi_0^{(0)}$}
\label{fini}
\end{figure}

We assessed the approach to the target state along the time by designing a 10 ps pulse having a cut-off frequency $\omega_{\rm max}=20$ THz for three values of $Z$ (Fig. \ref{fini}).
Fig. \ref{fini} shows the result of applying OCT to initialize the DQD device. The yield reaches a high value after about $10^3$ iterations of the optimization procedure, although typically $10^4$ iterations have been used in the calculations. 
The Fourier spectrum of the optimized pulses show only a few relevant 
frequencies giving smoothly oscillating fields experimentally realizable. 
These initialization pulses have different characteristics depending on the magnitude of the impurity charge. 
For the system with $Z=0.1$, the electron population of the 
ground state is gradually transferred to the first excited state until 
approximately a half of the pulse length, when both occupations reach 
about $70\%$ and $30\%$, respectively. In the second half of the pulse, the electron population is again restored to the ground state.  
Only the two lowest states are involved in the transition $\Psi_0\rightarrow \Psi_0^{(0)}$ because, for small $Z$, both ground states are rather similar, and the pulse frequency is mainly determined by the energy difference between the ground and first excited states of $H^{(Z)}$.

\begin{figure}\centering
\includegraphics[scale=0.75]{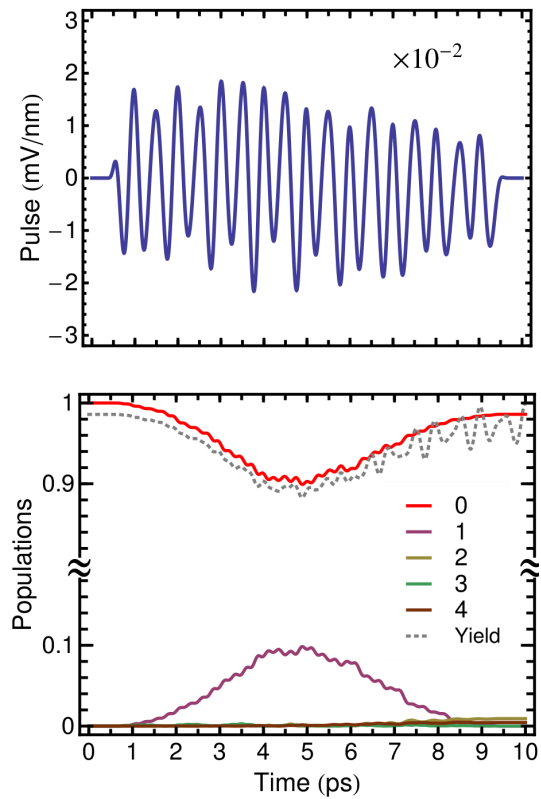}
\caption{(Color online) Upper panel: pulse of the 
initialization as a function of time for $F=1.3\times10^{-4} $mV$^2$/nm$^{2}$ for the 
double quantum dot with a Coulomb charge $Z=0.3$ placed in the middle of the 
segment of line joining the centers of the dots. Application of this pulse 
prepares the system to be used as a qubit, evolving the actual ground state 
of the system with an impurity $\psi_1^{(Z=0.3)}$ to the ground state of the 
system having no impurity $\psi_1^{(0)}$. Lower panel: evolution of the 
different states as a function of time.}
\label{papini}
\end{figure}

Fig. \ref{papini} shows the form of the optimized 10 ps pulse (upper panel) and the time dependence of the level occupations for the low lying states for $Z=0.3$ resulting from its application (lower panel). 
The pulse, although quite simple, is not monochromatic; the most relevant frequencies in its spectral composition are shifted downwards due to the 
level mixing between the DQD and the charged Coulomb ion, which have a smaller energy separation.
The  three dominant frequencies $\omega_{ij}$ can be identified as related to transitions between the low-lying levels $\Psi_i\rightarrow \Psi_j$, with the main contribution coming from $\omega_{01}\simeq 14$ THz and minor ones from   $\omega_{12}\simeq 0.13$ THz and  $\omega_{24}\simeq 4.6$ THz. 
The lower panel of Fig. \ref{papini} shows the time variation of the level occupation for the first five states during application of the pulse.
The ground state occupation $n_0$  decreases less than $10\%$ by the mid of the pulse duration, transferring population to the first excited state and, to less extent, to the second and fourth excited states. At the end of the pulse, the population $n_0$ remains high (close to $95\%$) while the rest 
populates the state $\Psi_3^{Z}$ and $\Psi_5^{Z}$.
In a wide range of cases studied, the pulses obtained from the OCT procedure for initialization of the qubit, under conditions of low fluence and frequency cut-off compatible with
current experimental capabilities, have been found to be rather simple, fast  and with fidelities higher than 99,9\%. 
\subsection{Operation of the qubit}
\label{op}

\begin{figure}\centering
\includegraphics[scale=0.55]{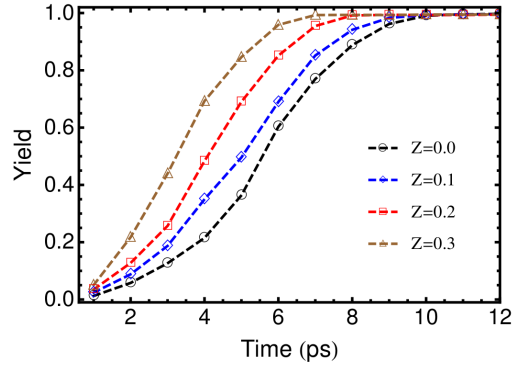}
\caption{\label{f5} (Color online) Yield of the operation pulse of fluence 
$F =1.3\times10^{-3}$ mV$^2$/nm$^2$ as a function of time for the double quantum dot 
with a Coulomb charge Z (for $Z=0.0$ (black circles), $Z=0.1$ (blue squares), 
$Z=0.2$ (red squares) and $Z=0.3$ (brown triangles)) placed in the middle 
of the segment
of line joining the centers of the dots. Application of this
pulse produces the transition between the localized ground state of the DQD 
without Coulomb impurity and the state where both electrons are localized at 
the right dot.}
\end{figure}

We address now the problem of producing the transition of interest for operating the qubit, $\Psi_0^{(0)}\rightarrow \Psi_{\rm RR}^{(0)}$, and how it is affected by the presence of the Coulomb charge.
Fig.  \ref{f5} shows the yield for the transition from the state $\Psi_0^{(0)}$, obtained with the initialization pulse discussed above, for $Z=0.1$, 0.2 and 0.3, together with the $Z=0.0$ case for the sake of comparison. The pulses were optimized to induce fast transitions (10 ps duration), yet they still hold high fidelity with the desired target state.
Interestingly,  Fig.  \ref{f5} shows that the system, in presence of the charge, evolves to the target state faster than the clean QDs. A yield of 90\% can be reached in around 7 ps for the $Z=0.3$ DQD, which is 2 ps shorter than for the clean system.
\begin{figure}\centering
\includegraphics[scale=0.75]{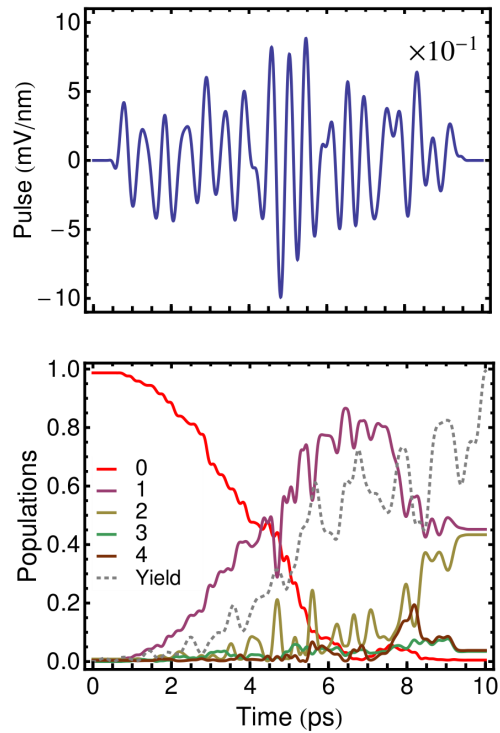}
\caption{\label{f6} (Color online) Upper panel: pulse of the                 
operation as a function of time for $F=1.3\times 10^{-3}$  mV$^2$/nm$^2$ for the
double quantum dot with a Coulomb charge $Z=0.3$ placed in the middle of the
segment of line joining the centers of the dots. Lower panel: evolution of 
the different states as a function of time.
}
\end{figure}

The optimal pulse for targeting the localized state $\Psi_{\rm RR}^{(0)}$, after the initialization pulse, is shown at the upper panel in Fig. \ref{f6}. The resulting yield and the variations of the occupation of the lower-lying states is depicted in the lower panel of Fig. \ref{f6}. As shown, the transition mainly involves the three lowest states of the DQD, namely, the ground state $\Psi_0 $, and first and second excited states, $\Psi_1 $ and $\Psi_2 $. The third and forth excited states $\Psi_3$ and $\Psi_4$ also gives, to a less extent, some smaller contribution during the second half of the pulse. The yield reaches a value $\gtrsim99.9$\% after 10 ps, with a steadily increasing behaviour although with oscillations of $\sim 10$\% during the second half of the pulse (i.e., between 5 and 10 ps). The examination of the time dependence of the state occupations allows us to explain how the target state is built and this high fidelity is reached.
During the firsts 4 ps, the population of the ground state is transferred almost exclusively to the state $|1\rangle$. In the next 4 ps (from 4 to 8 ps.) the population of the ground state $|0\rangle$ continues decreasing monotonically but part of the electronic charge is also transferred to the state $|2\rangle$. As a remarkable feature of the figure, the occupations of the states $|1\rangle$ and $|2\rangle$ show complementary peaks and dips of oscillations mounted on a smooth variation. Peaks of one curve occurs at  the dips of the other, entailing that part of the charge is oscillating between states $|1\rangle \leftrightarrow |2\rangle$. Finally, during the last 2 ps. the states $|1\rangle$ and $|2\rangle$ approach to be nearly equally populated in order to reach the target state. Nevertheless, there is a small but observable difference between them due to the occupation of higher excited states.

\begin{figure*}\centering

\hspace*{1cm} \includegraphics[scale=0.47]{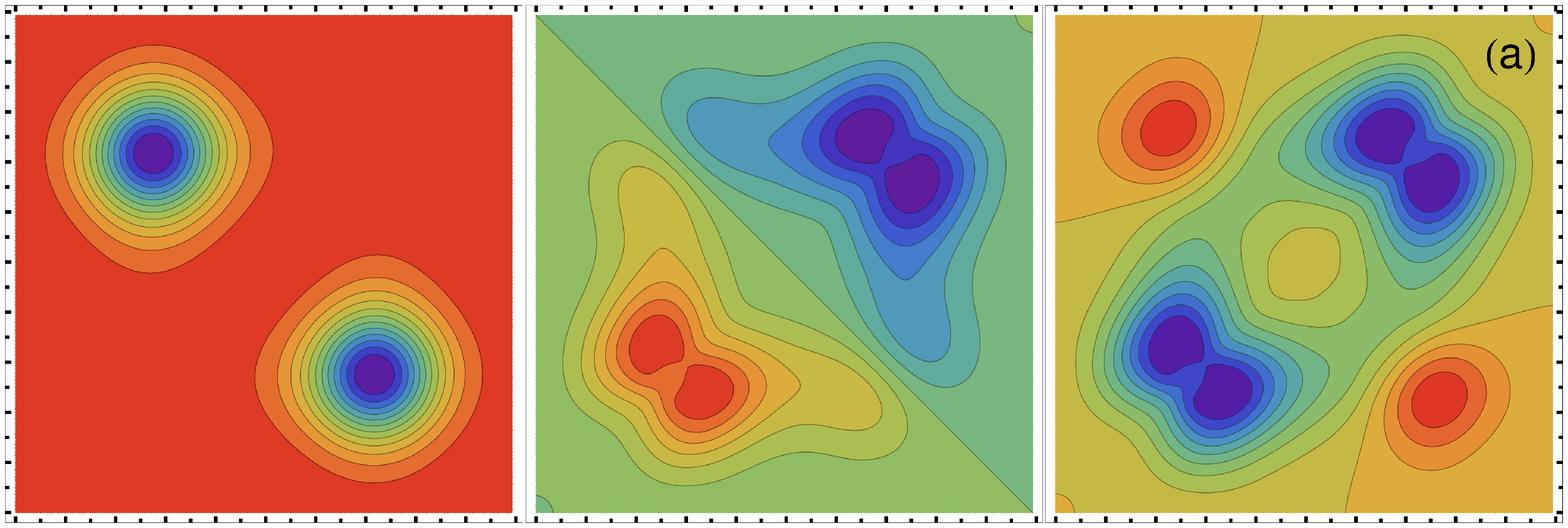} \\
\hspace*{1cm} \includegraphics[scale=0.47]{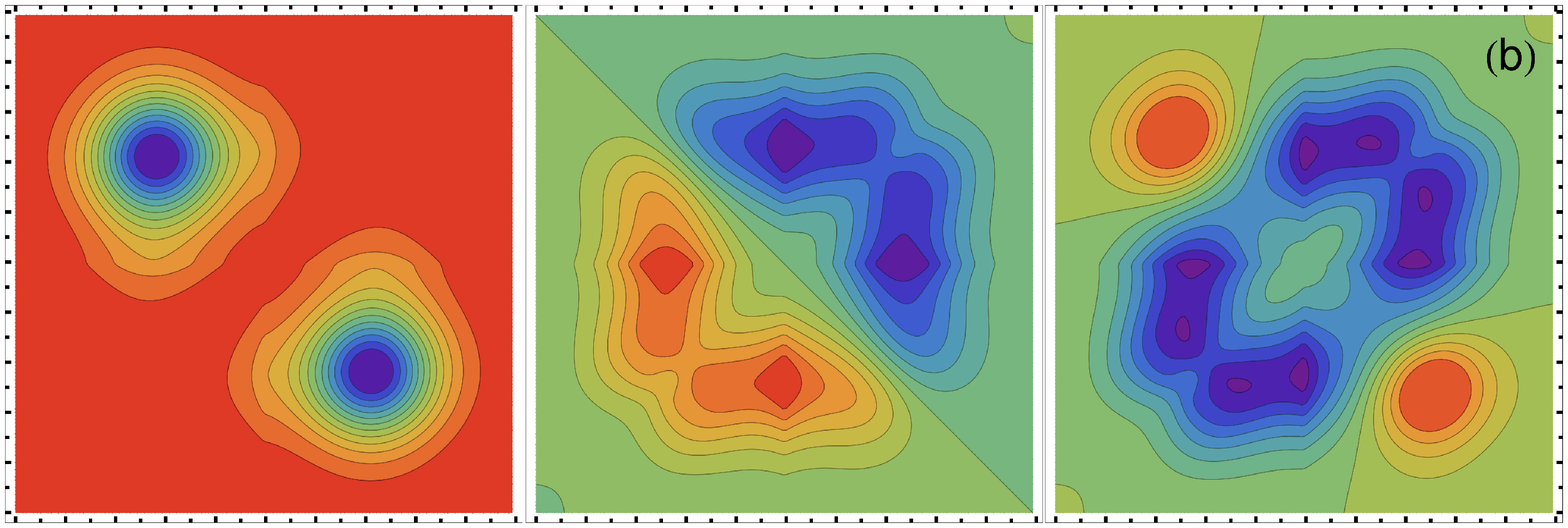} 

  \hspace{0cm} \includegraphics[scale=0.5]{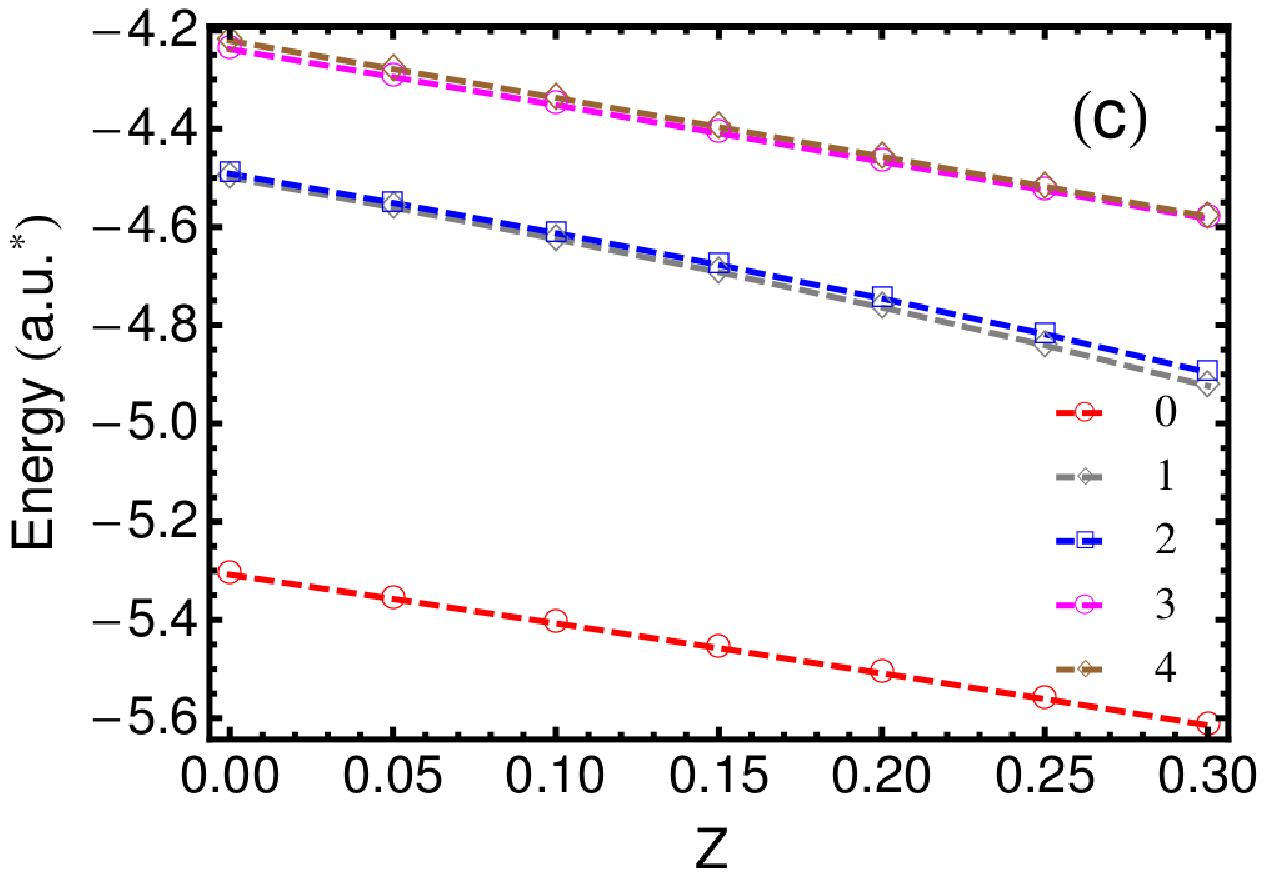} \hspace*{-0.7cm}
 \hspace{0cm} \includegraphics[scale=0.5]{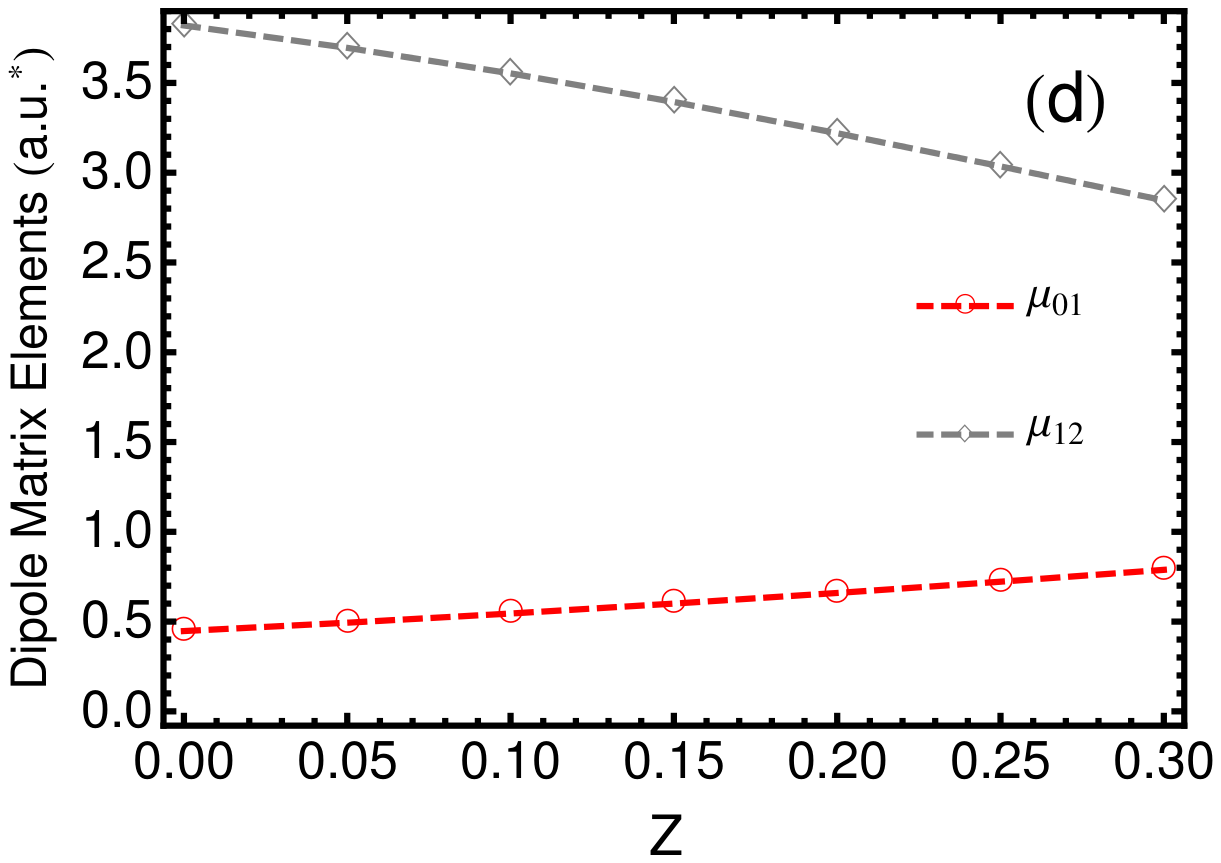}

\caption{\label{low lying states}  The calculated electronic structure of the double quantum dot. (a)-(b): Contour plot picture of the ground and two firsts excited states wave functions along the interdot line of the DQD without (a) and with (b) a Coulomb impurity charge $Z=0.3$ as a function of the $x_1$ and $x_2$ electron coordinates. Approximate expressions of the states are given by Eqs. (\ref{0})-(\ref{2}). (c) Variation of the low-lying electronic energies as a function of the charge $Z$, (d) Variation of the dipole moments $\mu_{01}$ and $\mu_{12}$ as a function of $Z$.}
\end{figure*}
The dynamical process of reaching the target state described above, and the influence of the impurity charge on the controllability, can be understood from an analysis of the electronic structure of our system, as sketched in Fig. \ref{low lying states}. 
It shows the two electron wave function along the $x$ axis joining both dots, $\Psi_i(x_1,0;x_2,0)$, as a contour plot of the variables ($x_1,x_2$), for the three lowest energy levels of the DQD without impurities (Fig. \ref{low lying states}a) and in the presence of an impurity of charge $Z=0.3$ (Fig. \ref{low lying states}b).
Positive and negative $x_i$ coordinates refer to positions of electron $i$ close to right and left wells, respectively. When the two-electron state corresponds to a situation where, spatially, each electron is in a different well, the wave function have large values along the $x_1=-x_2$ diagonal ($\searrow$) and represents a delocalized two-electron state. On the other hand, high values around the diagonal $x_1=x_2$ ($\nearrow$) entails for double occupation, i.e., when both electrons are in one of the wells.  
Therefore, irrespective of the presence or not of the Coulomb charge,  Figs. \ref{low lying states}a and \ref{low lying states}b show that the ground state $|0\rangle$ is a two-electron state describing electrons delocalized at different wells. Excited states $|1\rangle$ and $|2\rangle$, on the other hand, are mainly along the $\nearrow$ direction, i.e., they represent double occupation of the wells. State $|1\rangle$ has a nodal line along direction $\searrow$ while $|2\rangle$ has not, meaning that $|1\rangle$ and $|2\rangle$ have ungerade and gerade symmetry under inversion through the interdot center $x=0$.
The changes in the states from Fig. \ref{low lying states}a to Fig. \ref{low lying states}b, due to the charge $Z$, are apparent; states with $Z\ne 0$ have a noticeable contribution from the impurity location $x_1=x_2=0$. 
More information, to be discussed below, is provided in panels (c) and (d); they show the dependence of the electronic energy of the five lowest states (Fig. \ref{low lying states}c) and the two matrix elements of the dipole operator $X= x_1+x_2$ that give rise to the most relevant  transitions (Fig. \ref{low lying states}d), as a function of the Coulomb charge $Z$, respectively.

Further insight can be gained by approximating the system by a two-sites Hubbard model having hopping $w$,  on-site Coulomb repulsion $U$, and one orbital per site $\varphi_A$ (A=L, R). Assuming zero on-site energies for both wells and strong repulsion ($w\ll U$), its three lowest (non-normalized) singlet states and energies are
\begin{eqnarray}
\Psi_0 &=&     \Psi^g_S   +{\cal O}\left(
\frac{w}{U}\right)  \Psi^g_D  , 
\hspace{1cm} E_0  \approx -\frac{4w^2}{U} \label{0} \\
\Psi_1 &=&  \Psi^u_D   , 
\hspace{3.7cm} E_1 = U \label{1} \\
\Psi_2 &=&   \Psi^g_D  + {\cal O} \left(
\frac{w}{U}\right)  \Psi^g_S  , 
\hspace{1cm} E_2 \approx U +\frac{4w^2}{U} \label{2} 
\end{eqnarray}
The subscripts S and D stand for single or double occupation of the on-site orbitals $\varphi_{L,R}$, and  the superscripts $u$ and $g$ refer to the inversion symmetry (gerade or ungerade) with respect to the interdot center.
In the strongly correlated regime, $E_1$ and $E_2$ are quasi-degenerates, as in our CI calculations (Fig. \ref{low lying states}c). In such regime, our target state can be approximated as  $\Psi_{\rm RR} =(\Psi^g_D -\Psi^u_D)/2 \approx (\Psi_2 -\Psi_1)/2$.
Due to the point symmetry, dipole matrix elements $\mu_{ij}=\langle \Psi_i|X|\Psi_j\rangle$ are such that $\mu_{02}=0$, but $\mu_{01} \ne 0 \ne \mu_{12}$. Hence, transitions $0\leftrightarrow 1 \leftrightarrow 2$ are allowed, but  $0\leftrightarrow 2$ is not, as it is actually the case in our CI calculations. Within this model, $\mu_{01}=(4w/U)x_{LL}$ and $\mu_{12}=2x_{LL}$, with $x_{LL}=\int \varphi^2_L({\bf r})x d{\bf r}$ being the matrix element calculated in terms of the orbital centered at the left well $\varphi_L$. Therefore, from this approximate model a relation $\mu_{01}/\mu_{12}=2w/U$ is expected, with a transition probability lower for $0\leftrightarrow1$ than for $1\leftrightarrow2$. The corresponding resonant frequencies are in the relation $\omega_{12}/\omega_{01}\approx 4w^2/U^2\ll 1$.
The dynamics of our system can be thought in terms of this three-levels Hubbard model as follows: starting form the ground state $|0\rangle$ [eq. (\ref{0})], the external electric field induces transitions  $0\leftrightarrow1$ increasing the occupation of the first excited state, while $|2\rangle$ remains empty  because $0\leftrightarrow2$ is forbidden. After some time, part of the population of $|1\rangle$ is transferred to $|2\rangle$ due to the large $\mu_{12}$. These processes need to be excited with frequencies $\omega_{01}$ and $\omega_{12}$. The spectral composition of our optimally designed pulse of length $T$ shows two important contributions: one at $\omega_{01}\approx 14$ THz and other at a low frequency, which cannot be resolved because is less than $2\pi/T$, but could be related to $\omega_{12}$. Since our target state $|{\rm RR}\rangle$ requires to populate both $|1\rangle$ and $|2\rangle$, the processes continues until both become evenly populated. The presence, in our CI calculations, of the 
state $|4\rangle$ and higher levels (not included in the approximate model) having a non vanishing dipole moment matrix element $\mu_{24}$, produce some leakage, giving rise to the population of $|4\rangle$ observed in our calculations  (Fig. \ref{f6}).

To some extent, the effect of the low charge impurities can also be understood from the approximate Hubbard model. It can be shown that the effect of the addition of a one-electron operator to the Hubbard Hamiltonian (such as the Coulomb potential of the charged impurity or the electric field of the controlling laser) is accounted for by a change in the hopping parameter $w\rightarrow w'=w+\delta w$, with $\delta w \sim -Z$ for the Coulomb charge one-electron operator. 
The dipole moment operator $X=x_1+x_2$ has matrix elements $\langle \Psi_S^g|X|\Psi_D^u\rangle=0$ but $\langle \Psi_D^g|X|\Psi_D^u\rangle\ne 0$. The relative contribution of $|\Psi_D^g\rangle$ increases with $w$ (and therefore with $Z$) in $|0\rangle$ but decreases in $|2\rangle$ [eqs. (\ref{0}) and (\ref{2})] while $|1\rangle$ is independent of $w$. Hence $\mu_{01}$ increases, while $\mu_{12}$ decreases,  linearly with the impurity charge $Z$. The energy differences $\hbar \omega_{01}$ and $\hbar \omega_{12}$ do not change (at first order in $Z$) because all three energy eigenvalues share the same dependence. In our CI calculations, the probability of dipole transitions $0\leftrightarrow 1$ increases at a lower rate than the decreasing of the one for transitions $1\leftrightarrow 2$ (Fig. \ref{f6}d). As a consequence, if the pulse designed for the DQD without impurity, containing frequencies $\omega_{01}$ and $\omega_{12}$, is applied to the system doped with a charged impurity, the whole processes is 
slower and, at the end of the pulse duration, the resulting final state have a lower fidelity. 

In spite of the usefulness of the approximate Hubbard model for interpreting the results, one should be warned that the detailed electronic structure of the DQD becomes more and more relevant as higher values of fluence of the field are considered because of the increasing influence of the higher energy levels.

\begin{figure}\centering
\includegraphics[scale=0.6]{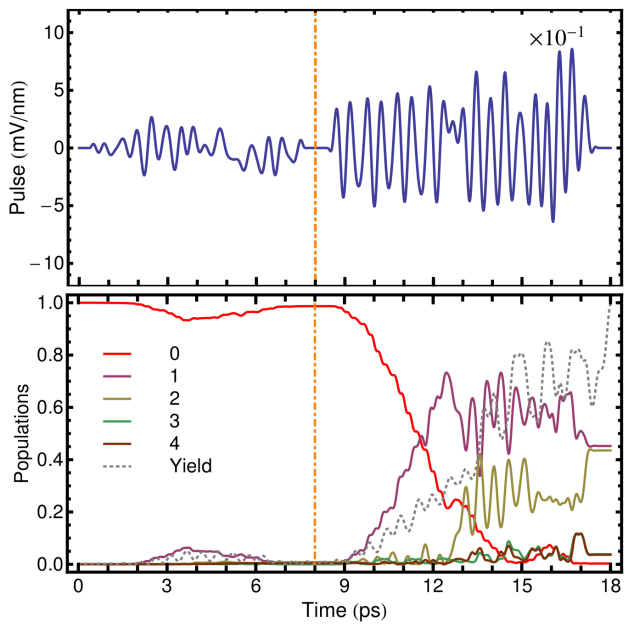}
\caption{\label{applied pulse} (Color online) Initialization and operation of the charge qubit in a double quantum dot with a Coulomb impurity.  The fluence of the laser pulse in the initialization step is $F =1.3\times10^{-4}$ mV$^2$/nm$^2$ whereas in the operation step it is of one higher order. Upper panel: Complete pulse designed in order to initializate and operate the doped device ($Z=0.3$). Lower panel: Population of the five lowest energy states during the transition $\Psi_0\rightarrow \Psi_0^{(0)}\rightarrow \Psi_{\rm RR}^{(0)}$.}
\end{figure}

The complete pulse resulting from the application of the proposed protocol and the time evolution of the level occupation of the system with $Z=0.3$ is depicted in Fig. \ref{applied pulse}. The vertical line at $t=8$ ps separates the two two steps of the process, i.e., 8 ps. for initialization and, then, 10 ps for operation of the qubit. Although in a device running tasks for information processing, the qubit transition will have to be run many times, the initialization step would be required just once; thus the whole process is not strongly affected by the impurity.

\section{Conclusions}
\label{con}

We have studied the efficiency of OCT based pulses suitable to produce transitions between localized and delocalized states of a double quantum dot device, with or without unintentional impurities. Those transitions give rise to changes of the electron charge in the individual dots, which can be detected, thus experimentally realizing a charge qubit. The resulting fast and high-fidelity tailored pulses are able to operate the qubit in times of the order of 10 ps, shorter than the decoherence time in clean samples, but their fidelity deteriorates heavily when even small Coulomb charges are present in the system. 
Therefore, we proposed and assessed the performance of applying a two-step protocol, by firstly initializing the electronic states in the ground state of the system without impurities, such that it compensates the changes in the electronic structure suffered by the DQD, introduced by the Coulomb charge. The second pulse operates the qubit as if it were impurity-free. Since both steps are designed in terms of the real electronic structure of the charged qubit, we have also analyzed the influence of the charge $Z$ on the electronic states of the double quantum dots. We have found that the complete two-step protocol involves mainly the lowest lying energy levels and remains fast enough to drive the states to the desired target state with a high fidelity, compatible with the requirements for its use in information processing tasks.

\section*{Acknowledgements}
We would like to acknowledge CONICET (PIP 112-201101-00981), SGCyT(UNNE) 
and FONCyT (PICT-2012-2866) 
for partial financial support of this project.

\end{document}